\documentclass[twocolumn,pra,showpacs,superscriptaddress]{revtex4}
\usepackage{amssymb}
\usepackage{amsmath}
\usepackage{graphicx}
\usepackage{subfigure}
\usepackage{natbib}
\usepackage{epsfig}
\usepackage{amsfonts}
\usepackage{mathrsfs}
\usepackage{CJK}
\usepackage[toc,page,title,titletoc,header]{appendix}

\begin{document}

\title{Quantum dynamical speedup in correlated noisy channels}

\author{Kai Xu}
 \affiliation{Key Laboratory of Micro-Nano Measurement-Manipulation and Physics (Ministry of
Education), School of Physics, Beihang University,
Xueyuan Road No. 37, Beijing 100191, China}

\author{Guo-Feng Zhang}
\email{gf1978zhang@buaa.edu.cn}
 \affiliation{Key Laboratory of Micro-Nano Measurement-Manipulation and Physics (Ministry of
Education), School of Physics, Beihang University,
Xueyuan Road No. 37, Beijing 100191, China}

\author{Wu-Ming Liu}
 \affiliation{Beijing National Laboratory for Condensed Matter Physics, Institute of Physics, Chinese Academy of Sciences, Beijing 100190, China}
 \affiliation{School of Physical Sciences, University of Chinese Academy of Sciences, Beijing 100190, China}
 \affiliation{Songshan Lake Materials Laboratory, Dongguan, Guangdong 523808, China}

\date{\today}
\begin{abstract}
The maximal evolution speed of a quantum system can be represented by quantum speed limit time (QSLT).
We investigate QSLT of a two-qubit system passing through a correlated channel (amplitude damping, phase damping, and depolarizing).
By adjusting the correlation parameter of channel and the initial entanglement,
a method to accelerate the evolution speed of the system for some specific channels is proposed.
It is shown that, in amplitude damping channel and depolarizing channel,
QSLT may be shortened in some cases by increasing correlation parameter of the channel and initial entanglement, which are in sharp contrast to phase damping channel.
In particular, under depolarizing
channels, the transition from no-speedup evolution to speedup evolution for the system can be realized by changing correlation strength of the channel.

\end{abstract}

\maketitle
\setlength{\parskip}{0.4\baselineskip}
\section{\textbf{{Introduction}}}

How to drive the initial state of a quantum system to the target state in
the minimum evolution time is a fundamental and important issue of quantum
physics \cite{Lloyd1,Giovannetti,Batle,Casas,F,Yung,Caneva1,Hegerfeldt1,Lloyd3}.
The quantum speed limit time (QSLT) defines the minimum evolution time between two
given states of a quantum system. It sets the maximal rate with which quantum information
can be processed \cite{Lloyd1}, the maximal rate with which quantum information can be
communicated \cite{Bekenstein}, the maximal rate of quantum entropy production \cite{Deffner},
the shortest time-scale for quantum optimal control algorithms to converge
\cite{Mukherjee,Hegerfeldt,Hegerfeldt2,Avinadav}, and determines the spectral
form factor \cite{Campo A}. \textbf{Furthermore, quantum speed limit time is closely related to
fields such as quantum metrology and quantum computation \cite{Childs,Giovanetti,Lloyd6}.
For instance, a practical problem that is beyond the reach of classical computation has been solved by quantum simulation with quantum speedup \cite{Childs}.}
Originally, for a closed system with unitary evolution,
QSLT is obtained by unifying Mandelstam-Tamm (MT) type bound and Margolus-Levitin (ML)
type bound \cite{Mandelstam,Fleming,Margolus,Levitin}. Since a quantum system inevitably
interacts with environment, the bounds of evolution time including both Mandelstam-Tamm
(MT) and Margolus-Levitin (ML) types focused on an open system with nonunitary dynamics
process have also been formulated \cite{Campaioli,Mirkin,Taddei,Zhang,Lidar,Xu}. More specifically,
three independent approaches of how to quantify the maximal quantum speed of systems in
noisy channels have been proposed. Taddei et al. \cite{Taddei} found an expression in terms of
the quantum Fisher information to quantify quantum speed of system in the typical noisy
channels. Del Campo et al. \cite{Egusquiza} bounded the rate of change of the relative purity
to provide the speed of evolution under an open-system dynamics. And Deffner and Lutz \cite{Lutz}
derived a Margolus-Levitin-type bound on the minimum evolution time of an arbitrary driven open
quantum system. These results have caused the interest of some further research about quantum speed
limit.

Recently, some remarkable progress has been made in analyzing the effects of environment
on open quantum systems. Here people focus on two aspects of the environment's impact on open systems:
transition between quantum states and loss of phase coherence induced by dissipative or
pure decoherence environments \cite{Suter,Schlosshauer,Laine,Piilo,Vacchini,Cai,Cai1,Cai2}. For example, \textbf{the dynamical speedup of a
quantum system in a non-equilibrium environment has been investigated \cite{Cai}.}
The decoherence speed limit
for spin-deformed bosonic model and the impacts of nonlinear environment have been considered
in pure phase-damping channel \cite{Dehdashti}.
The non-Markovianity and the formation of the system-environment
bound states have been proven to accelerate the speed of evolution of the quantum states by studying the dynamics of a
dissipative two-level system \cite{An,Fan,Sun,Meng,Zhangyj4,Wang0}. Furthermore, the speedup evolution of a given N-qubit
entangled state may occur by controlling the number of independent amplitude-damping channels \cite{Han}.

The above studies are concentrated on the sequence of qubits passing through a uncorrelated channel by neglecting the
correlations between multiple uses of quantum channels. However, in the laboratory, the effects of correlation on the evolution
process of qubits are inevitable. For instance, in quantum information processors, especially in
solid-state implementations, qubits may be so closely spaced that the same environmental degree of freedom will interact
jointly with several of them (even if they are not nearest
neighbors) leading to cross talks and correlations in the noise \cite{Duan,Zhou,Lupo}. So it
is very meaningful to study how the correlation strength of channel affects the dynamic evolution of quantum system.

In this paper, we consider a model with classical correlations between two consecutive applications of channel.
We use the quantum speed limit time to evaluate the speed of quantum state evolution and consider the three common noise sources:
the amplitude damping channel, phase damping channel and depolarizing channel. In addition, we analyze in detail the
effects of consecutive applications of the noisy channel and the entanglement of the initial state on dynamical evolution process of the two-qubit system. \textbf{The results show that, in the amplitude damping channel and the depolarizing channel, QSLT can be shortened by increasing correlation strength of the channel and initial entanglement, which are exactly the opposite of the phase damping channel. Furthermore, it is worth emphasizing that, in depolarizing channel, the quantum system experiences a process from no-speedup to speedup evolution by changing correlation strength of the channel.}

\textbf{This paper is organized as follows. In Sec. II we briefly introduce the correlated quantum channel and quantum speed limit time.
Sec. III discusses the influence of correlation strength of the channel and the initial entanglement on the dynamical speedup of
the quantum evolution for different noisy models in three subsection. The conclusions drawn from the present study are given in Sec. IV.}

\section{\textbf{Correlated quantum channel and QSLT}}

In this section, we first review two different types of quantum channels including memoryless channels and memory channels.
For memoryless channels (uncorrelated channels), the quantum channels act identically and independently on each of the quantum system.
More specifically, quantum channel $\varepsilon$ for N consecutive uses obey $\varepsilon_{N}=\varepsilon^{\otimes N}$.
However, in the reality, the correlations between consecutive uses of the channel on a set of quantum
systems may exist, so that the channel acts dependently on each channel input, $\varepsilon_{N} \neq \varepsilon^{\otimes N}$.
Such channels are called memory channels (correlated channels).

For the sake of simplicity, we consider two uses of quantum channels.
Given the initial state $\rho_{0}$ of the system, the output state can be obtained

\begin{eqnarray}
\rho=\sum_{i_{1} i_{2} } E_{i_{1} i_{2} } \rho_{0} E_{i_{1} i_{2} }^{\dagger}  \label{01},
\end{eqnarray}
where $E_{i_{1} i_{2} }=\sqrt{p_{i_{1} i_{2}}} B_{i_{1}} \otimes B_{i_{2}}$
are the Kraus operators of the channel which satisfy the completeness relationship, and $\sum_{i_{1} i_{2}} p_{i_{1} i_{2}}=1$. Here
$p_{i_{1} i_{2}}$ represents the joint probability. For uncorrelated channels, these operations $B_{i_{1}} \otimes B_{i_{2}}$
are independent and therefore lead to $p_{i_{1} i_{2}}=p_{i_{1}} p_{i_{2}}$. However, for a correlated channel,
these operations are time-correlated. Macchiavello and Palma \cite{Palma} proposed such a model, for which the joint probability is given by

\begin{equation}
p_{i_{1} i_{2}}=(1-\mu) p_{i_{1}} p_{i_{2}}+\mu p_{i_{1}} \delta_{i_{1} i_{2}},
\end{equation}
where $\mu \in[0,1]$ represents the degree of classical correlation
in the performance of channel. For $\mu=0$, this model depicts independent applications of channel,
while for $\mu=1$, the applications of the channel become fully correlated.
Then the Kraus operators for two consecutive uses of a channel with partial correlation are

\begin{equation}
E_{i_{1} i_{2}}=\sqrt{p_{i_{1}}\left[(1-\mu) p_{i_{2}}+\mu \delta_{i_{1} i_{2}}\right]} B_{i_{1}} \otimes B_{i_{2}} \label{01}.
\end{equation}

With the above description, we will focus on the noisy channel (i.e., amplitude damping, phase damping, and depolarizing channels) for two consecutive uses.
Based on the Kraus operator approach, for any initial state $\rho_{0}$, the final state of the system in correlated noise channel can be given by \cite{Palma,Yeo,Macchiavello}

\begin{equation}
\begin{aligned}
\rho &=(1-\mu) \sum_{i_{1} i_{2}} E_{i_{1} i_{2}} \rho_{0} E_{i_{1} i_{2}}^{\dagger}+\mu \sum_{k} E_{k k} \rho_{0} E_{k k}^{\dagger}\\
&=(1-\mu) \varepsilon_{un}+\mu \varepsilon_{co},
\end{aligned}
\end{equation}
where $\varepsilon_{un}$ represents the uncorrelated channel and $\varepsilon_{co}$ stands for correlated channel.
Eq. (4) implies that the same operation is applied to two qubits with probability $\mu$, whereas different operations are applied to two qubits with probability $1-\mu$.

Next, to study the speed of dynamical evolution of the system, we
need to start with the definition of QSLT for an open
quantum system. QSLT can effectually define the bound of minimal evolution
time for arbitrary initial states, and be helpful to analyze the maximal evolution speed of an open quantum
system. Deffner et al. \cite{Lutz} derived a unified lower bound of the quantum speed
limit time which is determined by an initial state $\rho_{0}$ and its target state $\rho_{\tau_{D}}$.
With the help of von Neumann trace inequality and
Cauchy-Schwarz inequality, QSLT can be written

\begin{equation}
\tau_{Q S L}=\max \left\{\frac{1}{\Lambda_{\tau_{D}}^{1}}, \frac{1}{\Lambda_{\tau_{D}}^{2}}, \frac{1}{\Lambda_{\tau_{D}}^{\infty}}\right\} \sin ^{2}\left[\mathbf{B} \left(\rho_{0}, \rho_{\tau_{D}}\right)\right], \label{03}
\end{equation}
with $\Lambda_{\tau_{D}}^{l}=\tau_{D}^{-1} \int_{0}^{\tau_{D}}\left\|\dot{\rho}_{t}\right\|_{l} d t$, and $\|A\|=\left(\sigma_{1}^{l}+\dots+\sigma_{n}^{l}\right)^{1 / l}$
denotes the Schatten $l$-norm, $\sigma_{1}, \sigma_{2}, \cdots, \sigma_{n}$ are the singular
values of $A$, $\tau_{D}$ denotes the driving time. $\mathbf{B}\left[\rho_{0}, \rho_{\tau_{D}}\right]=\arccos \sqrt{\left\langle\phi_{0}\left|\rho_{\tau_{D}}\right| \phi_{0}\right\rangle}$ denotes the Bures angle between the initial and target states of the quantum
system. Based on the operator norm ($l$=$\infty$, that is $\Lambda_{\tau_{D}}^{\infty}=\tau_{D}^{-1} \int_{0}^{\tau_{D}} \max \left[\sigma_{1}, \sigma_{2}, \cdots, \sigma_{n}\right] d t$) of the nonunitary generator, the ML-type bound provides the sharpest bound on QSLT \cite{Lutz}. Therefore, for the initial pure state, we use this ML-type
bound to demonstrate QSLT of the dynamics evolution from an initial state $\rho_{0}$ and its target state $\rho_{\tau_{D}}$. However, Eq. (5) is not feasible for mixed initial states.
Fortunately, based on the relative purity along with von Neumann trace
inequality and Cauchy-Schwarz inequality, a unified lower bound on QSLT including both MT and
ML types has been derived for arbitrary initial states in open quantum systems \cite{Zhang}, which reads

\begin{equation}
\tau_{Q S L}=\max \left\{\frac{1}{\overline{\sum_{i=1}^{n} \sigma_{i} \rho_{i}}}, \frac{1}{\overline{\sqrt{\sum_{i=1}^{n} \sigma_{i}^{2}}}}\right\} *\left|f_{\tau+\tau_{\mathrm{D}}}-1\right| \operatorname{Tr}\left(\rho_{\tau}^{2}\right) \label{03},
\end{equation}
where $\overline{X}=\tau_{\mathrm{D}}^{-1} \int_{\tau}^{\tau+\tau_{\mathrm{D}}} X \mathrm{d} \mathrm{\tau}$, $\sigma_{i}$ and $\rho_{i}$ are respectively the singular values of $\dot{\rho}_{t}$ and
 $\rho_{\tau}$, and $f\left(\tau+\tau_{D}\right)=\operatorname{tr}\left[\rho_{\tau+\tau_{D}} \rho_{\tau}\right] / \operatorname{tr}\left(\rho_{\tau}^{2}\right)$
denotes the relative purity between the initial state $\rho_{\tau}$  and the final state $\rho_{\tau+\tau_{D}}$  with the driving time $\tau_{D}$. According to Ref. \cite{Liu},
$\tau_{Q S L} / \tau=1$ means the quantum system evolution is already along the fastest path and possesses no potential capacity for further quantum speedup.
While for the case $\tau_{Q S L} / \tau<1$, the speedup evolution of the quantum system may occur and the much shorter $\tau_{Q S L} / \tau$, the greater
the capacity for potential speedup will be.

\section{\textbf{QSLT in correlated noisy channel}}

We firstly use the ML-type bound to calculate QSLT of the dynamics evolution
from an initial state $\rho_{0}$ to a final state $\rho_{\tau}$ by fixing an actual evolution time $\tau$.
And then, since the correlations between multiple uses of quantum channels are inevitable in experiments, we mainly discuss how the
existence of the correlation can be more favorable to the acceleration of the quantum state than
the uncorrelated channels. Furthermore, the effect of initial entanglement on the speed of evolution of quantum states is also considered.
Below we mainly study the above problems from three basic channels: amplitude damping, phase damping and depolarizing damping.

\subsection{\textbf{Quantum speedup in correlated amplitude damping channel}}

As we all know, the amplitude damping channel describes relaxation processes,
such as spontaneous emission of an atom, in which the system decays from the excited state $| 1 \rangle$ to the ground state $|0\rangle $.
The Kraus operators for a single qubit are given by

\begin{equation}
B_{i_{1}}=\left( \begin{array}{cc}{1} & {0} \\ {0} & {\sqrt{P}}\end{array}\right), \quad B_{i_{2}}=\left( \begin{array}{cc}{0} & {\sqrt{1-P}} \\ {0} & {0}\end{array}\right) \label{01},
\end{equation}
where $P=e^{-\Gamma t}$ is the decay of the excited population, and $\Gamma$ is the dissipation rate. If two qubits are considered to pass the uncorrelated amplitude damping channel,
the Kraus operators are defined as the following form
\begin{equation}
E_{i_{1}i_{2}}=B_{i_{1}} \otimes B_{i_{2}},(i_{1}, i_{2}=0,1) \label{01}.
\end{equation}
A full-memory amplitude damping channel was introduced in Ref. \cite{Yeo}. And the Kraus operators $E_{kk}$ is given by
\begin{equation}
E_{00}=\left( \begin{array}{cccc}{1} & {0} & {0} & {0} \\ {0} & {1} & {0} & {0} \\ {0} & {0} & {1} & {0} \\ {0} & {0} & {0} & {\sqrt{P}}\end{array}\right), \quad \\
E_{11}=\left( \begin{array}{cccc}{0} & {0} & {0} & {\sqrt{1-P}} \\ {0} & {0} & {0} & {0} \\ {0} & {0} & {0} & {0} \\ {0} & {0} & {0} & {0}\end{array}\right).
\end{equation}

In this work, we consider the initial state $\rho_{0}=|\Phi\rangle\langle\Phi|$, where $\Phi=\alpha|00\rangle+\beta|11\rangle$ corresponds to the Bell-like states with $0\leq\alpha\leq1$.
According to Eq. (4), in the correlated amplitude damped channel, the evolved density matrix of the two-qubit system, whose elements
in the standard computational basis $\{|1\rangle =|11\rangle,|2\rangle = | 10\rangle,|3\rangle=|01\rangle,|4\rangle =|00\rangle \}$ are

\begin{equation}
\begin{aligned} \rho_{11} &=\alpha^{2}-(-1+P) \beta^{2}(1+P(-1+\mu)), \\
\rho_{22} &=(-1+P) P \beta^{2}(-1+\mu), \\
\rho_{33} &=(-1+P) P \beta^{2}(-1+\mu), \\
\rho_{44} &=P \beta^{2}(P+\mu-P \mu), \\
\rho_{14} &=\rho_{41}=-P \alpha \beta(-1+\mu)+\sqrt{P} \alpha \beta \mu.
\end{aligned}
\end{equation}
Based on Eq. (10), we have
$\sin ^{2}\left[\mathbf{B}\left(\rho_{0}, \rho_{\tau}\right)\right]$=$|\mathrm{Tr}(\rho_{0}\rho_{\tau})-1|$=$|-1+\alpha^{4}+P \beta^{4}(P+\mu-P \mu)+\alpha^{2} \beta^{2}\left(1-P^{2}(-1+\mu)+2 \sqrt{P} \mu-P \mu\right)|$.
Then our task in the following is to calculate the singular values of $\dot{\rho_{t}}$ and find out the largest
singular value $\sigma_{\max }=\left\|\dot{\rho}_{t}\right\|_{\infty}$.
After a simple calculation, the singular values $\sigma_{i}$ are
$\sigma_{1}$=$\sigma_{2}$=$|(-1+\mu)\beta^{2}(-1+2P)||\dot{P}|$, $\sigma_{3}$=$|(-1+\mu)\beta^{2}(-1+2P)+(\beta/{2})|\mu \alpha P^{\frac{1}{2}}+2(1-\mu)\alpha|\sqrt{1+4\beta^{2}/(\mu\alpha P^{\frac{1}{2}}+2(1-\mu)\alpha)^{2}}||\dot{P}|$, $\sigma_{4}$=$|(-1+\mu)\beta^{2}(-1+2P)-(\beta/{2})|\mu \alpha P^{\frac{1}{2}}+2(1-\mu)\alpha|\sqrt{1+4\beta^{2}/(\mu\alpha P^{\frac{1}{2}}+2(1-\mu)\alpha)^{2}}||\dot{P}|$.
And $\sigma_{max}=|(-1+\mu)\beta^{2}(-1+2P)|+(\beta/{2})|\mu \alpha P^{\frac{1}{2}}+2(1-\mu)\alpha|\sqrt{1+4\beta^{2}/(\mu\alpha P^{\frac{1}{2}}+2(1-\mu)\alpha)^{2}}|\dot{P}|$. Therefore, in the correlated amplitude damping channel, Eq. (5) can be written as $\tau_{QSL}/\tau=|-1+\alpha^{4}+P_{\tau} \beta^{4}(P_{\tau}+\mu-P_{\tau} \mu)+\alpha^{2} \beta^{2}\left(1-P_{\tau}^{2}(-1+\mu)+2 \sqrt{P_{\tau}} \mu-P_{\tau} \mu\right)|/{\int_{0}^{\tau}\sigma_{max}dt}$, wi-th $P_{\tau}$ means the excited population of the final state $\rho_{\tau}$.
It is clearly to find that the QSLT of the two-qubit state is evaluated as a function of the correlation parameter $\mu$
and the initial entanglement $(\alpha,\beta)$.

\begin{figure}[tbh]
\includegraphics*[bb=16 27 246 376,width=8cm, clip]{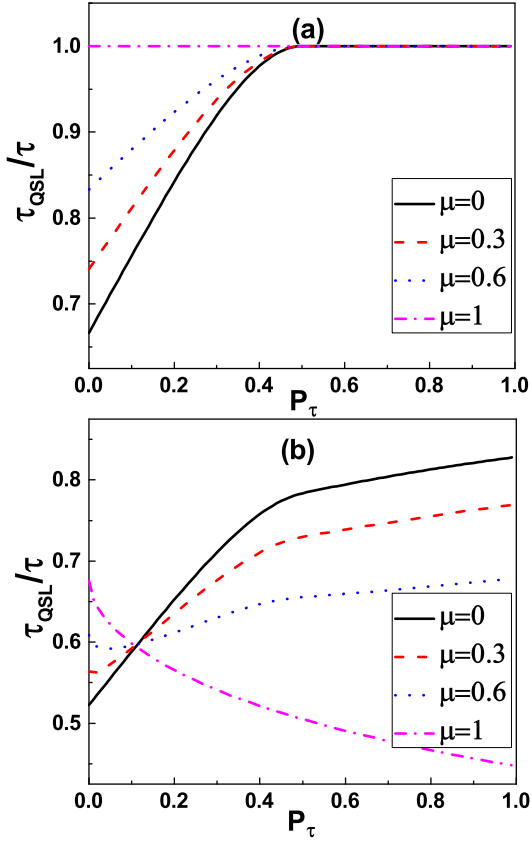}
\caption{(Color online) QSLT for a given two-qubit state,
quantified by $\tau_{QSL}/\tau$ as a function of the
excited population $P_{\tau}$ of the final state.
(a) for the initial unentangled state $\alpha=0, \beta=1$; (b) for the initial entangled state $\alpha=\sqrt{2}/{2}, \beta=\sqrt{2}/{2}$.}
\end{figure}

\begin{figure}[tbh]
\includegraphics*[bb=104 179 352 382,width=8cm, clip]{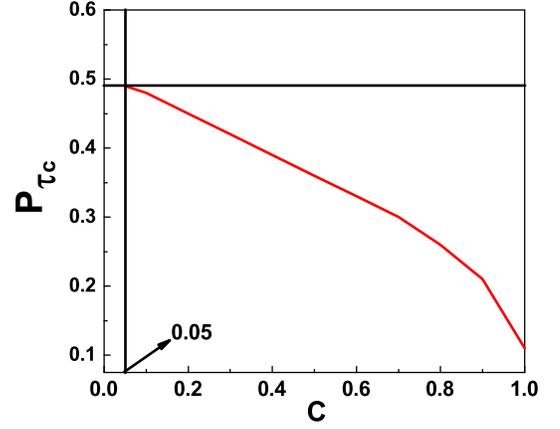}
\caption{(Color online) Dependence of the critical population $P_{\tau_{c}}$ of excited state of the final state on the initial entanglement $C$.}
\end{figure}

By confirming an actual evolution time $\tau$, the influences of the correlation strength of channel $\mu$ and the excited population $P_{\tau}$
on QSLT are depicted in Fig. 1. The entanglement of the system can be characterized by
Wootter's concurrence. For the initial two-qubit state $\rho_{0}$, the concurrence can be obtained $C=2|\alpha \beta|$. In Fig. 1(a),
we first consider the case where the initial state is prepared in an unentangled state. By fixing $P_{\tau}$, QSLT increases or
does not change as $\mu$ increases. That is to say, when the initial state is unentangled, the increase of $\mu$ inhibits the speedup evolution of the system.
However, for the initial entangled state in Fig. 1(b), the larger $\mu$ can lead to the greater potential speedup of the evolution process at a certain region $[P_{\tau_{c}},1]$
($P_{\tau_{c}}$ means a certain critical value of $P_{\tau}$). Here $P_{\tau_{c}}$ is related to initial entanglement $C$, as shown in Fig. 2.
More specifically, by fixing $\mu=0, 0.3, 0.6, 1$ in Fig. 2, in the case $0<C<0.05$, there is no $P_{\tau_{c}}$, suggesting that the increase of $\mu$ can not accelerate the evolution of the quantum state.
While $C>0.05$, the critical value $P_{\tau_{c}}$ is a monotonic decreasing function of $C$. This implies that the specific region $[P_{\tau_{c}},1]$ that accelerates the
evolution of the quantum state by increasing $\mu$ is broader with increasing $C$.
Combining the above analysis, we find that when $P_{\tau}>P_{\tau_{c}}$ and the initial entanglement $C>0.05$ are satisfied, the existence of the correlation parameter $\mu$ of the channel increases the speed of evolution of the system compared to the uncorrelated channel $\mu=0$.

\begin{figure}[tbh]
\includegraphics*[bb=88 140 334 353,width=8cm, clip]{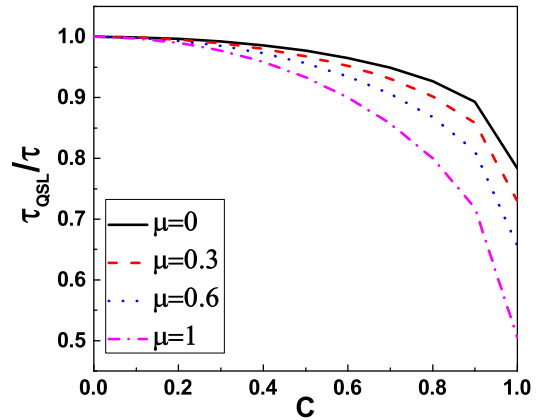}
\caption{(Color online) QSLT for a given two-qubit state,
quantified by $\tau_{QSL}/\tau$ as a function of the
entanglement $C = 2|\alpha\beta|$ of the initially prepared state.
Parameter is chosen as $P_{\tau}=0.5$.}
\end{figure}

In the following, to more intuitively explain the effect of the initial entanglement $C$ on QSLT,
we fix $P_{\tau}=0.5>P_{\tau_{c}}^{max}$ in Fig. 3. Clearly, in the cases $\mu=0,0.3,0.6,1$,
a remarkable dynamical crossover from no-speedup evolution to speedup evolution for the system can occur at a certain
critical initial entanglement $C_{c}$. When $C<C_{c}$, the system has no-speedup behavior, and then the capacity for potential
speedup of the system increases with increasing $C$.
In addition, it's valuable to point out that the value of this critical initial entanglement $C_{c}$ is independent of $\mu$.
Finally, it should be emphasized that, in the correlated amplitude-damped channel,
the larger initial entanglement can lead to the
greater potential speedup of the evolution process.

\subsection{\textbf{Quantum speedup in correlated phase damping channel}}

The phase damping channel describe a decoherencing process without exchanging energy with the environment.
The Kraus operators of a single qubit in the phase damped channel are represented as the Pauli operators
$\sigma_{0}=I$ and $\sigma_{3}$. Suppose that two qubits pass through memoryless dephasing channel,
the Kraus operators can be written as

\begin{equation}
E_{i_{1} i_{2}}=\sqrt{p_{i_{1}} p_{i_{2}}} \sigma_{i_{1}} \otimes \sigma_{i_{2}}\label{01},
\end{equation}
where $i_{1}, i_{2}$=(0, 3), $p_{0}=(1+p)/{2}$, $p_{3}=(1-p)/{2}$, and $p=\exp (-\gamma t)$.
For the partial correlated dephasing channel, the Kraus
operators $E_{k k}$ are given as

\begin{equation}
E_{k k}=\sqrt{p_{k}} \sigma_{k} \otimes \sigma_{k}\label{01}, \quad(k=0,3).
\end{equation}
Substituting Eqs. (11) and (12) into Eq. (4), the density matrix elements of the two qubits in the
correlated phase damped channel can be expressed in the following form

\begin{equation}
\begin{aligned}
\rho_{11} &=\alpha^{2},\\
\rho_{44} &=\beta^{2}, \\
\rho_{14} &=\rho_{41}=\alpha \beta\left(1- \left(1-p^{2})(1-\mu\right)\right).&
\end{aligned}
\end{equation}

Only the two nondiagonal terms of this two qubits density matrix decay,
the evolution under the correlated phase damping channel is easier to analyze. In what follows
we mainly focus on the influence of correlation strength of channel and initial entanglement on QSLT.
Based on Eq. (13), $\sin ^{2}\left[\mathbf{B}\left(\rho_{0}, \rho_{\tau}\right)\right]$=$2 |(1-p^{2}) \alpha^{2} \beta^{2}(-1+\mu)|$.
The singular values are $\sigma_{1}=\sigma_{2}=|2p\alpha\beta(-1+\mu)||\dot{p}|$, $\sigma_{3}=\sigma_{4}=0$.
Therefore, Eq. (5) can be simplified as $\tau_{QSL}/{\tau}=2 |(1-p_{\tau}^{2}) \alpha^{2} \beta^{2}(-1+\mu)|/{\int_{0}^{\tau}|2p\alpha\beta(-1+\mu)||\dot{p}|dt}$=$(C/{2})|1-p_{\tau}^{2}|/{\int_{0}^{\tau}p|\dot{p}|dt}$.
We find that QSLT increases with the increase of initial entanglement $C$, which show that enhancing initial entanglement will suppress the capacity for potential speedup of the quantum state.
Furthermore, another meaningful result can be acquired from the above QSLT expression: in the correlated damped channel, QSLT of the dynamics evolution from $\rho_{0}$ to $\rho_{\tau}$
does not depend on the correlation parameter $\mu$ for a given two-qubit pure initial state. Then one might wonder whether this is true for the arbitrary initial time parameter of the system.

\begin{figure}[tbh]
\includegraphics*[bb=98 181 356 392,width=8cm, clip]{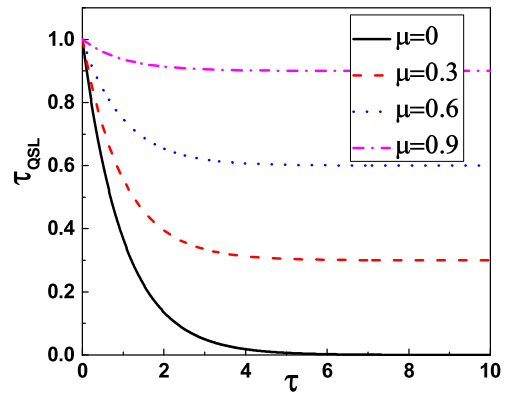}
\caption{(Color online) $\tau_{QSL}$ as a function of the initial time parameter $\tau$.
Parameters are chosen as $\alpha=\beta=\sqrt{2}/{2}$, $\gamma=1/{2}$, $\tau_{D}=1$.}
\end{figure}

To resolve this problem, we explore the effects of the correlation sterength of channel $\mu$ on QSLT in the whole dynamical process. For $\gamma=1/{2}$,
$\tau_{QSL}=2e^{-\tau}\alpha\beta(1-\mu)+2\alpha\beta\mu$.
With this, it is easy to found that $\tau_{QSL}$ is related to the dephasing rate, initial entanglement£¬ and the correlation parameters.
Figure 4 shows the results of our analysis for $\tau_{QSL}$ as a function of the initial time parameter $\tau$ by choosing different correlation strength $\mu$.
We discover that, for the same driving time
$\tau_{D}=1$, the larger correlation strength of channel can decrease the speed of evolution of a quantum system, and
thus demand the longer $\tau_{QSL}$. Besides, by considering the relatively larger initial time parameter $\tau$, the quantum speed limit time
can be rewritten as $\tau_{QSL}=2\alpha\beta\mu$. Therefore, when $\alpha=\beta=\sqrt{2}/{2}$ is fixed in Fig. 4, the value of $\tau_{QSL}$ will stabilize around $\mu$
after a finite limited time. Finally, it is worth emphasizing that, in the correlated phase-damped channel, the increase of $\mu$ does not have a favorable effect
on the reduction of quantum speed limit time of the system.

\subsection{\textbf{Quantum speedup in correlated depolarizing channel}}

The depolarizing noise is the quantum operation that depolarizes the state into a completely mixed state.
The Kraus operators for a single qubit are $B_{i}=\sqrt{p_{i}} \sigma_{i}$ $(i=0,1,2,3)$, where
$p_{0}=(1+p)/{2}$, $p_{1}=p_{2}=p_{3}=(1-p)/{6}$, and $p=\exp (-\gamma t)$. Assume the time interval between channel successive
applications on the two qubits is infinitely small, the uncorrelated depolarizing channel model can be applicable.
At this point, the Kraus operators $E_{i_{1} i_{2}}$ can be written as
\begin{equation}
E_{i_{1} i_{2}}=\sqrt{p_{i_{1}} p_{i_{2}}} \sigma_{i_{1}} \otimes \sigma_{i_{2}}, \quad(i_{1}, i_{2}=0,1,2,3).
\end{equation}
Here we will consider the case of two consecutive uses of depolarizing channel, the Kraus operator $E_{kk}$ can be given
as

\begin{equation}
E_{k k}=\sqrt{p_{k}} \sigma_{k} \otimes \sigma_{k}, \quad(k=0,1,2,3).
\end{equation}
According to Eq. (4), the density matrix elements of the two qubits are

\begin{equation}
\begin{aligned}
\rho_{11}&=-\frac{1}{9}(2+p) \alpha^{2}(-2+p(-1+\mu)-\mu)-\frac{1}{9}(-1+p) \beta^{2}\\
\times&(1+p(-1+\mu)+2 \mu),\\
\rho_{22} &=\rho_{33} =\frac{1}{9}\left(-2+p+p^{2}\right)(-1+\mu),& \\
\rho_{44}&=-\frac{1}{9}(-1+p) \alpha^{2}(1+p(-1+\mu)+2 \mu)+\frac{1}{9} \beta^{2}\\
\times&((2+p)^{2}-\left(-2+p+p^{2}\right) \mu),\\
\rho_{14}&=\rho_{41}=\frac{1}{9} \alpha \beta(1-4 p(1+p)(-1+\mu)+8 \mu).&
\end{aligned}
\end{equation}
Based on Eq. (16), we can clearly find, $\sin ^{2}\left[\mathbf{B}\left(\rho_{0}, \rho_{\tau}\right)\right]=|(-1+p)\left(5+p\left(-1-8 \beta^{2}+8 \beta^{4}\right)(-1+\mu)\right)/{9}-(-1+p)2 \mu/{9}+(-1+p) 4 \beta^{2}\left(-1+\beta^{2}\right)(-1+4 \mu)/{9}|.$ The singular values are $\sigma_{1}=\sigma_{2}=|(1+2p)(1-\mu)||\dot{p}|/{9}$, $\sigma_{3}=\{|(1+2p)\times(1-\mu)+(1+2p)\times(1-\mu)\sqrt{16\alpha^{2}\beta^{2}+9(1-4\alpha^{2}\beta^{2})/{(\mu-1)^{2}(2p+1)^{2}}})|\times|\dot{p}|\}/9$, and $\sigma_{4}=\{|(1+2p)\times(1-\mu)-(1+2p)\times(1-\mu)\sqrt{16\alpha^{2}\beta^{2}+9(1-4\alpha^{2}\beta^{2})/{(\mu-1)^{2}(2p+1)^{2}}})|\times|\dot{p}|\}/{9}$. Clearly, $\sigma_{3}$ is the largest singular value. Then the quantum speed limit time can be obtained
$\tau_{QSL}/\tau=\sin ^{2}\left[\mathbf{B}\left(\rho_{0}, \rho_{\tau}\right)\right]/\int_{0}^{\tau}\sigma_{3}dt$.

\begin{figure}[tbh]
\includegraphics*[bb=37 19 269 376,width=8cm, clip]{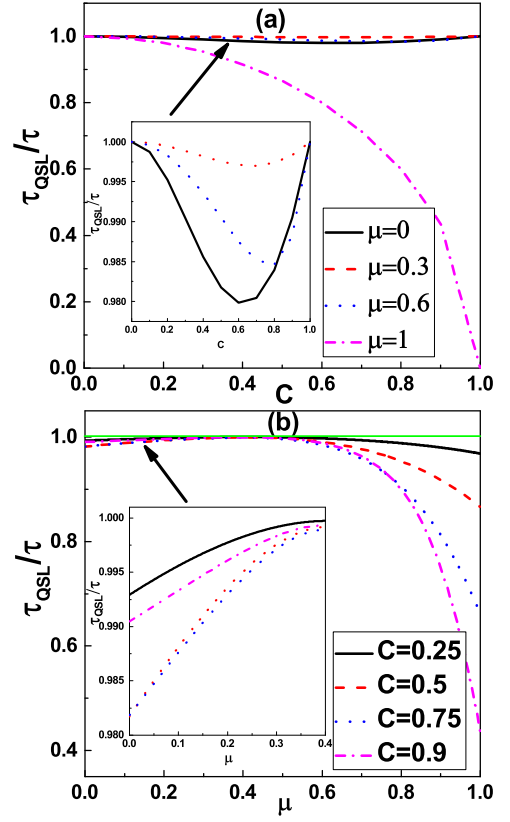}
\caption{(Color online) QSLT for a given two-qubit state,
quantified by $\tau_{QSL}/\tau$ as a function of the parameters for the
entanglement $C = 2|\alpha\beta|$ of the initially prepared state and the correlation parameter $\mu$. (a) and (b) $p_{\tau}=0.5$.}
\end{figure}

Next, we also focus on the effects of initial entanglement and the correlation strength of channel
on QSLT. Firstly, when the channels are partially correlated or not correlated (i.e., $\mu=0, 0.3, 0.6$),
$\tau_{QSL}/\tau$ changes nonmonotonically with the increase of initial entanglement $C$, as shown by the inset
of Fig. 5(a). Specifically, $\tau_{QSL}/\tau$ is reduced from one to the minimum and then restored to one.
This means that the maximum potential speedup of evolution process can be achieved by preparing appropriate initial entanglement $C$.
Differently, when the channels are fully correlated (i.e., $\mu=1$) in Fig. 5(a), the quantum speed limit time can be written as $\tau_{QSL}/\tau=\sqrt{1-C^{2}}$, implying the larger initial entanglement can lead to the
greater potential speedup of the evolution process. Besides, we find that the quantum system does not evolve (i.e., $\rho_{0}=\rho_{\tau}$) when the initial state is the maximum entangled state or the separable state.
Therefore, to better study the change of $\tau_{QSL}/\tau$ with $\mu$, the initial entanglement state ($0<C<1$) is chosen in Fig. 5(b).
We can clearly see that the accelerated evolution of quantum states can appear in a specific region $[\mu^{critical},1]$.
And it is worth noting that, for initial entangled state ($0<C<1$),
the maximal capacity for potential speedup of a given two-qubit state would be reached in fully correlated depolarizing channel ($\mu=1$).

\section{\textbf{{Conclusion}}}

In conclusion, we have explored dynamical evolution process of two qubits when they pass through a correlated noisy channel.
By using the quantum speed limit time to define the speedup evolutional process, we discuss the influence of initial
entanglement and the correlation strength of channel on the speed of evolution of the quantum state in decoherence channels
(amplitude damping channel, phase damping channel and depolarizing channel). We find that, in the correlated phase damping channel,
enhancing the correlation strength of channel and initial entanglement could inhibit the speed of evolution of
quantum systems. However, in correlated amplitude damping channel, for excited population $P_{\tau}$  and initial entanglement $C$ ,
there are thresholds beyond which the correlation strength of channel would increase the evolution speed of the system compared with uncorrelated channel.
Furthermore, for the initial entangled state ($0<C<1$) in the correlated depolarizing channel,
the transition from no-speedup phase to speedup phase of the system can be realized by increasing the correlation parameter of channel.
Therefore, in some specific channels, the existence of correlation parameters can play a beneficial role in accelerating the evolution of the system.
Finally, since any physical process can be represented as a quantum channel mapping an initial state to a final state, the decoherence channel model (amplitude-damping channel, phase-damping channel, depolarizing channel) considered here is rather general, and the obtained results are hoped to be of guiding significance in
understanding the dynamic evolution process of the system immersed in real environments.

\section{\textbf{{Acknowledgements}}}
This work was supported by NSFC under grants Nos. 11574022,
11434015, 61835013, 11611530676, KZ201610005011,
the National Key R\&D Program of China under grants Nos. 2016YFA0301500,
SPRPCAS under grants No. XDB01020300, XDB21030300. Also, we would like to thank Y. J. Zheng and X. J. Cai for useful discussions.

\end{document}